\documentclass[pra,twocolumn,superscriptaddress]{revtex4}
\usepackage{graphicx}
\usepackage{amsmath}
\usepackage{amssymb}
\usepackage{braket}
\usepackage[colorlinks=true,linkcolor=blue,citecolor=blue,urlcolor=blue]{hyperref}

\begin{document}

\title{Confinement-induced Resonance of Alkaline-earth-metal-like Atoms in Anisotropic Quasi-one-dimensional Traps}

\author{Qing Ji}
\affiliation{Department of Physics, Renmin University of China, Beijing 100872, China}
\author{Ren Zhang}
\email{renzhang@xjtu.edu.cn}
\affiliation{Department of Applied Physics, School of Science, Xi'an Jiaotong University, Shanxi 710049, China}
\author{Xiang Zhang}
\email{siang.zhang@ruc.edu.cn}
\affiliation{Department of Physics, Renmin University of China, Beijing 100872, China}
\author{Wei Zhang}
\email{wzhangl@ruc.edu.cn}
\affiliation{Department of Physics, Renmin University of China, Beijing 100872, China}
\affiliation{Beijing Key Laboratory of Opto-electronic Functional Materials and Micro-nano Devices, Renmin University of China,
Beijing 100872, China}

\date{\today}
\begin{abstract}
We study the confinement-induced resonance (CIR) of $^{173}$Yb atoms near an orbital Feshbach resonance in a quasi-one-dimensional tube with transversal anisotropy.  By solving the two-body scattering problem, we obtain the location of CIR for various anisotropy ratio and magnetic field. Our results show that the anisotropy of the trapping potential can serve as an additional knob to tune the location of CIR. In particular, one can shift the location of CIR to the region attainable in current experiment. We also study the energy spectrum of the system and analyze the properties of CIR from the perspective of bound states. We find that as the orbital Feshbach resonance acquires two nearly degenerate scattering channels, which in general have different threshold energies, CIR takes place when the closed channel bound state energy becomes degenerate with one of the thresholds.
\end{abstract}

\maketitle

\section{Introduction}
Confinement-induced resonance (CIR) plays a key role in the quantum simulation of 1D quantum models with cold atoms~\cite{oshanii98,oshanii03,CIRexp,Wei-Peng,Shiguo-Hui, saenz, schmelcher}, by means of which the effective interaction strength for a quantum atomic gas confined in an elongated quasi-1D geometry can be resonantly enhanced. As was firstly proposed by Olshanii~\cite{oshanii98}, the low-energy two-body scattering processes within a quasi-1D trapping potential can be well described by an effective 1D model, where the 1D interaction strength is intimately related to both the three-dimensional (3D) $s$-wave scattering length $a_{s}$ and the external confinement. Later, it is realized that this phenomenon can be understood in the framework of Feshbach resonance, where the ground state in the strongly confined transversal plane is regarded as the open channel, and the excited states as a whole assume the closed channel. The effective 1D interaction becomes resonant as the bound state in the closed channel degenerates with the open channel threshold~\cite{oshanii03}. Since in the dilute limit, the properties of the system are dominated by two-body processes, this observation paves the avenue towards the simulation of 1D quantum many-body Hamiltonian with tunable interaction in a quasi-1D configuration. With the aid of CIR, the Tonks-Girardeau gas and super Tonks-Girardeau gas have been explored experimentally in cold alkali-metal atoms~\cite{1dFermion,1dboson}.

The study of CIR has also been then extended to alkali-metal atoms in quasi-1D confinement with transversal anisotropy~\cite{CIRexp,Wei-Peng,Shiguo-Hui, saenz, schmelcher}. The presence of anisotropy can serve as another knob to tune the effective 1D interaction strength and the position of CIR~\cite{Wei-Peng,Shiguo-Hui, schmelcher}. Besides, the anharmonicity of the transversal trapping potential is suggested to be crucial to explain the splitting of resonances observed in experiments~\cite{saenz, CIRexp}. Besides, the inclusion of synthetic spin-orbit coupling can provide another tuning parameter that can alter the position of CIR~\cite{1D-zhangren, zhang-liu-15}.

Recently, the ultracold quantum gas of alkaline-earth-metal-like (AE) atoms has attracted great attention owing to its unique advantages in quantum simulation. In addition to the electronic ground state $^{1}S_{0}$, the AE atom has an extremely long-lived clock state $^{3}P_{0}$ whose single-atom lifetime can be as long as many seconds. The ground and clock state manifolds hence can serve as an additional degree of freedom, which is referred as {\it orbit}. For fermionic AE atoms, the nuclear spin is nonzero and decouples from the electronic degree of freedom in the ground and clock states. As the interatomic interaction is dominated by electronic wave functions, the nuclear spin degree of freedom supports an SU$(N)$ symmetry, where $N$ denotes the number of possible nuclear spin states. This property paves the route towards the investigation of fermionic systems with a plethora of SU($N$) symmetry in AE atoms~\cite{Kondo_alkali_earth,Takahashi,Ray,1dSUN-Fallani,cnyang,JILA-spin-ex,Munich-spin-ex,Florence-spin-ex,SUN-HM}.

Another distinctive feature of AE atoms is the existence of spin-exchange scattering between the ground state $^{1}S_{0}$ and clock state $^{3}P_{0}$~\cite{JILA-spin-ex,Munich-spin-ex,Florence-spin-ex}, which is not only crucial for the technique of orbital Feshbach resonance (OFR) to tune the interatomic interaction~\cite{ren1, ofrexp1, ofrexp2}, but also a essential ingredient for the simulation of Kondo physics~\cite{Kondo_alkali_earth,Nakagawa,ren,yanting,nagy}. It is then of great interest to find a way to enhance the spin-exchange interaction strength, and in turn the Kondo temperature to the extent achievable in current experiments. One possible method to achieve this goal is to tune the system through a CIR in a quasi-1D confinement.~\cite{ren,yanting} To one's delight, the resonant enhancement of spin-exchange scattering near CIR has been demonstrated in $^{173}$Yb atoms~\cite{spin-ex-exp}. In such experiment, the ground state and clock state atoms are trapped in isotropic quasi-1D tubes simultaneously by magic wave length lasers along the $x$- and $y$-directions, while the clock state atoms are further trapped by an external potential along the $z$-direction. Thus, the intensity of lasers along the transversal ($x$--$y$) and axial($z$) directions can be viewed as two knobs to control the location of CIR, which enriches the tool box for the simulation with AE atoms.

In the current manuscript, we investigate the two-body scattering problem of AE atoms in a quasi-1D confinement with transversal anisotropy, i.e., the confinement strengths along the $x$- and $y$-directions are in general different. By investigating the low-energy two-body scattering processes, we show that the transversal anisotropy of the trapping potential can serve as another tuning knob to manipulate the location of CIR. For alkali-metal atoms in quasi-1D confinement, CIR can be regarded as a Feshbach resonance where the transversal ground state and excited states as a whole assume the open and closed channels, respectively~\cite{oshanii03, Wei-Peng, 1D-zhangren} However, for AE atoms near an OFR, there are four atomic states involved and hence two scattering channels which are slightly shifted by Zeeman effect. As a result, by confining the AE atoms in a quasi-1D trap, the two channels would acquire different open channel threshold energies. A natural question would be which one of the thresholds should be intersected by the closed channel bound state energy when CIR occurs. By comparing with the solution of scattering states, we show that CIR takes place when the closed channel bound state energy is degenerate to only one of the threshold energies.

The remainder of the manuscript is organized as follows. In Sec.~\ref{sec:formalism}, we present the formalism of low-energy two-body scattering between two AE atoms near an OFR, first in a quasi-1D trap with transversal anisotropy, and then using an effective 1D model. By matching solutions of the two models, we obtain the effective interaction strength in the 1D model in Sec.~\ref{sec:interaction}. Then, we discuss in Sec.~\ref{sec:boundstate} the energy spectrum of the two-body bound state and analyze CIR from the perspective of bound state. Finally, we summarize in Sec.~\ref{sec:conclusion}.

\begin{figure}
\centering
\includegraphics[width=0.4\textwidth]{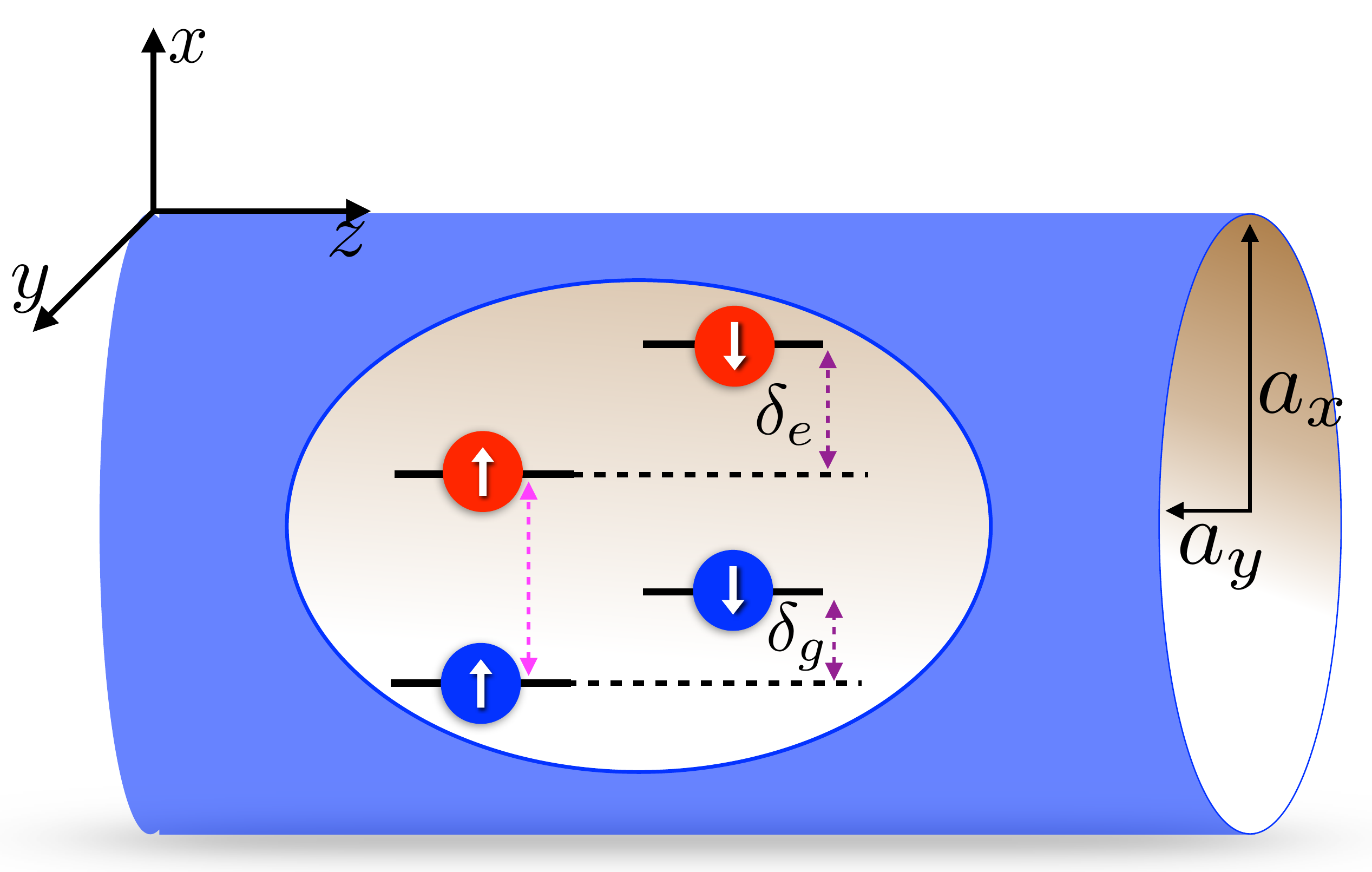}
\caption{Energy levels of alkaline-earth-metal-like atoms trapped in an anisotropic quasi-one-dimensional tube. The blue and red balls denote atoms in the ground state $^{1}S_{0}$ and clock state $^{3}P_{0}$, respectively. The up and down arrows denote different nuclear spin states. The trapping potential is produced by a laser with magic wave length which can trap the ground state and clock state simultaneously. $a_{x}$ and $a_{y}$ are the typical lengths of the trapping potential along the $x$- and $y$-direction, respectively, which can be tuned by changing the laser intensity and profile. The Zeeman energy difference between the $|g\uparrow;e\downarrow\rangle$ and $|g\downarrow;e\uparrow\rangle$ channels is $\delta \equiv \delta_{e}-\delta_{g}$, where $\delta_{g}$ and $\delta_{e}$ are the Zeeman energy of the ground state and clock state manifolds, respectively.}\label{cartoon}
\end{figure}

\section{Scattering states}
\label{sec:formalism}
The leading goal of dimension reduction from quasi-one to one dimension is to find a concise 1D effective model Hamiltonian which captures the low-energy physics of the original quasi-1D Hamiltonian by integrating out the degrees of freedom with high energy. For a Fermi gas at low temperature, the energy regime of interest is usually close to the Fermi surface, which lies within the low-energy regime for a dilute system. Thus, a natural scheme is to match the low-energy scattering amplitude close to the scattering threshold of quasi-1D system to that of a 1D model. In this section, we solve the scattering problem for AE atoms close to an OFR in both quasi-1D and 1D geometries, and determine the position of CIR from the resulting scattering amplitude.

\subsection{Quasi-one-dimension}
First we would like to introduce some basic properties of AE atoms.
Considering two AE atoms in the ground $^{1}S_0$ and clock $^{3}P_0$ states (denoted respectively by the orbital $|g\rangle$ and $|e\rangle$ states) with different nuclear spin states (denoted by pseudo-spin $|\uparrow\rangle$ and $|\downarrow\rangle$) confined in a quasi-one-dimensional tube as shown in Fig.~\ref{cartoon}, one can define two sets of basis $\{|+\rangle,|-\rangle\}$ and $\{|g\uparrow;e\downarrow\rangle,|g\downarrow;e\uparrow\rangle\}$, which are $|\pm\rangle\equiv(1/2)(|ge\rangle\pm |eg\rangle)(|\uparrow\downarrow\rangle\mp|\downarrow\uparrow\rangle)$ and $|gs;es'\rangle\equiv(1/\sqrt{2})(|gs\rangle|es'\rangle-|es'\rangle|gs\rangle)$ with $\{s, s'\}=\{\uparrow,\downarrow\}$. The two sets of basis are related to each other via the unitary transformation $|\pm\rangle=(1/\sqrt{2})(|g\uparrow;e\downarrow\rangle\mp|g\downarrow;e\uparrow\rangle)$. One of the salient features of the AE atoms is that the interaction Hamiltonian is diagonal in the basis of $\{|+\rangle,|-\rangle\}$, i.e., $\hat{V}=V_{+}({\bf r})|+\rangle\langle+|+V_{-}({\bf r})|-\rangle\langle-|$. The associated $s$-wave scattering length of $V_{\pm}({\bf r})$ are denoted by $a_{s\pm}$. For $^{173}$Yb atoms, $a_{s+}\approx1900a_{0}$ and $a_{s-}\approx219.5a_{0}$ with $a_{0}$ the Bohr's radius.
On the other hand, the free Hamiltonian is diagonal in the basis of $\{|g\uparrow;e\downarrow\rangle,|g\downarrow;e\uparrow\rangle\}$. Thus the total Hamiltonian can be written as
\begin{align}
\label{hamilton}
\hat{H}_{\rm Q1D}=&\left[-\frac{\hbar^2}{2\mu}\nabla_{\mathbf{r}}^2+\frac{\mu}{2} \omega_{x}^2x^2+\frac{\mu}{2}\omega _{y}^2y^2+V_{0}({\bf r})\right]\left({\cal P}_{\uparrow\downarrow}+{\cal P}_{\downarrow\uparrow}\right)\nonumber\\
&+\delta {\cal P}_{\uparrow\downarrow}+V_{1}({\bf r})({\cal S}_{\rm ex}+{\cal S}_{\rm ex}^{\dagger}),
\end{align}
where $\mu$ is the reduced mass of the two atoms, and $\omega_{x}$ and $\omega_{y}$ are the trapping frequencies along the $x$- and $y$-directions, respectively. To quantify the anisotropy strength of the tapping potential, we define the ratio of $\zeta=\omega_{x}/\omega_{y}$. The specific case of $\zeta=1$ means the trapping potential is isotropic in the transverse plane. The projection operators ${\cal P}_{ss'}$ and ${\cal S}_{\rm ex}$ are defined as ${\cal P}_{ss'}=|g s;e s'\rangle \langle g s;es'|$ and ${\cal S}_{\rm ex}=|g\uparrow;e\downarrow\rangle \langle g \downarrow;e\uparrow |$, respectively. The Zeeman energy difference between the $|g\uparrow;e\downarrow\rangle$ and $|g\downarrow;e\uparrow\rangle$ channels is denoted by $\delta=(m_{\uparrow}-m_{\downarrow})(\delta\mu)B$, where $(\delta\mu)=2\pi\hbar\times112$Hz/Gauss for $^{173}$Yb atoms. The interaction potential $V_{0}({\bf r})=(V_{+}({\bf r})+V_{-}({\bf r}))/2$ and $V_{1}({\bf r})=(V_{-}({\bf r})-V_{+}({\bf r}))/2$ denote the diagonal and non-diagonal parts of the interaction, respectively. In our calculation, we employ the following pseudo-potentials to simulate the involved van der Waals potentials,
\begin{align}
\label{HY}
V_{i=0,1}({\bf r})=\frac{2\pi\hbar^{2}a_{si}}{\mu}\delta({\bf r})\frac{\partial}{\partial r}r
\end{align}
with $a_{s0}=(a_{s+}+a_{s-})/2$ and $a_{s1}=(a_{s-}-a_{s+})/2$.

Now we consider the scattering problem in the low-energy limit. Specifically, the low-energy limit means that we assume the incident kinetic energy $\epsilon$ and the Zeeman energy $\delta$ is much smaller than the energy gap of the harmonic trap, i.e. $\{\epsilon,\delta\} \ll \{ \hbar \omega_{x}, \hbar \omega_{y} \}$. Hence, when the two atoms are faraway, the interaction potential decays to zero and the system is dominated by the free Hamiltonian. The two atoms thus reside in the ground state $|n_{x}=0,n_{y}=0\rangle$ of the transverse trapping potential with $n_x$ and $n_y$ the quantum numbers of the corresponding harmonic potentials. When the two atoms collide in a short distance, the interaction potential becomes important and couples the ground state to excited states of the trapping potential.

\begin{figure*}
\includegraphics[width=0.85\textwidth]{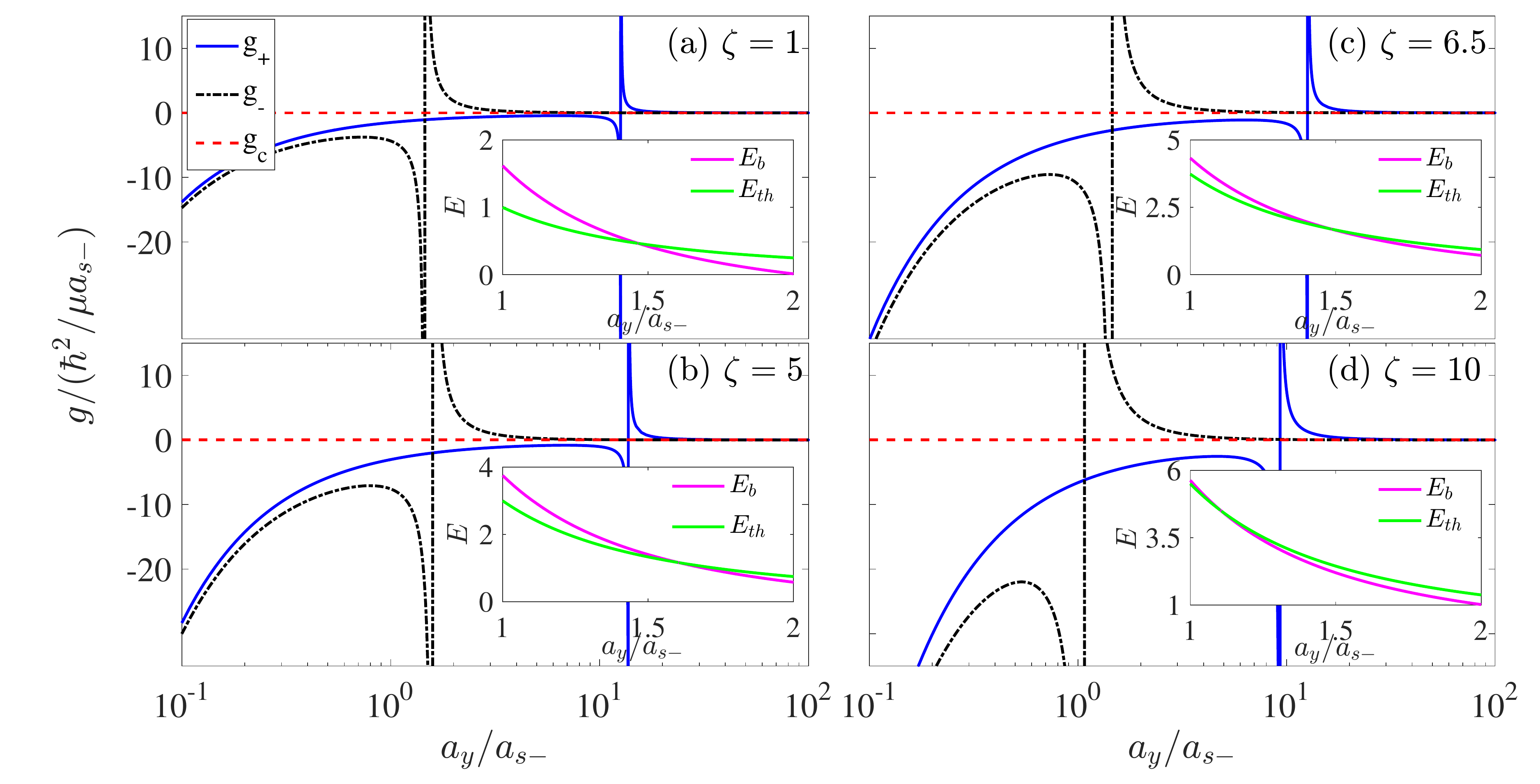}
\caption{Effective one-dimensional interaction $g$'s as functions of trapping length $a_{y}$. In our calculation, we take $^{173}$Yb atom as an example and use the following parameters $a_{s+}=1900a_{0}$ and $a_{s-}=219.5a_{0}$, $\omega_{x}=\zeta\omega_{y}$ and $\delta=0$. Note that $g_{c}$ is always zero since the $|+\rangle$ and $|-\rangle$ channels are decoupled in the absence of effective magnetic filed. The location of CIR can be manipulated by changing the anisotropy ratio. In each panel, the closed channel bound state energy $E_b$ and the open channel threshold $E_{th}$ are shown in the insets, with the energy unit $\hbar^2/\mu a_{s-}^2$. The position of the intersection between $E_b$ and $E_{th}$ coincides with the location of CIR determined by a scattering state analysis.}\label{g-ay}
\end{figure*}

The general scattering wave function associated with the Hamiltonian Eq.~(\ref{hamilton}) takes the following form
\begin{align}
|\Psi({\bf r})\rangle=\Psi^{(\uparrow\downarrow)}({\bf r})|g\uparrow;e\downarrow\rangle+\Psi^{(\downarrow\uparrow)}({\bf r})|g\downarrow;e\uparrow\rangle,\label{psir}
\end{align}
where the functions $\Psi^{(\uparrow\downarrow)}({\bf r})$ and $\Psi^{(\downarrow\uparrow)}({\bf r})$
are given by
\begin{align}
\Psi^{(\uparrow\downarrow)}({\bf r}) & =  \left[\alpha e^{ik^{(\uparrow\downarrow)}z}+f^{(\uparrow\downarrow)}{\rm e}^{ik^{(\uparrow\downarrow)}|z|}\right]\zeta^{1/4}\phi_{0}(\sqrt{\zeta}x)\phi_{0}(y)\nonumber\\&+\sum_{(n_x,n_y)}'B_{n_x,n_y}^{(\uparrow\downarrow)}{\rm e}^{-\kappa_{n_x,n_y}^{(\uparrow\downarrow)}|z|}\phi_{n_x}(\sqrt{\zeta}x)\phi_{n_y}(y);\label{psiud}\\
\Psi^{(\downarrow\uparrow)}({\bf r}) & =  \left[\beta e^{ik^{(\downarrow\uparrow)}z}+f^{(\downarrow\uparrow)}{\rm e}^{ik^{(\downarrow\uparrow)}|z|}\right]\zeta^{1/4}\phi_{0}(\sqrt{\zeta}x)\phi_{0}(y)\nonumber\\&+\sum_{(n_x,n_y)}'B_{n_x,n_y}^{(\downarrow\uparrow)}{\rm e}^{-\kappa_{n_x,n_y}^{(\downarrow\uparrow)}|z|}\phi_{n_x}(\sqrt{\zeta}x)\phi_{n_x}(y).\label{psidu}
\end{align}
Here, $\alpha$ and $\beta$ denote the possible amplitudes of the incident atoms in the $|g\uparrow;e\downarrow\rangle$ and $|g\downarrow;e\uparrow\rangle$ channels, respectively. Considering the zero-range pseudo-potential in which the scattering state with odd parity along the $z$-direction fails to feel the potential, we only focus on the even parity part such that the summation $\sum^\prime$ runs over all even combinations of $(n_x,n_y)$ except the ground state $n_x=n_y=0$. Besides, $\phi_{0}(x)$ is the ground state wave function of 1D harmonic oscillator, and the parameters $k^{(\uparrow\downarrow)}$,
$k^{(\downarrow\uparrow)}$, $\kappa_{n_x,n_y}^{(\uparrow\downarrow)}$
and $\kappa_{n_x,n_y}^{(\downarrow\uparrow)}$ are defined as
\begin{eqnarray}
k^{(\uparrow\downarrow)}&=&\sqrt{2\mu(\epsilon-\delta)/\hbar^{2}},
\nonumber \\
\kappa_{n_x,n_y}^{(\uparrow\downarrow)}&=&\sqrt{2\mu(n_x\hbar\omega_{x} +n_y\hbar\omega_{y} -\epsilon+\delta)/\hbar^{2}},
\nonumber \\
k^{(\downarrow\uparrow)}&=&\sqrt{2\mu\epsilon/\hbar^{2}},
\nonumber \\
\kappa_{n_x, n_y}^{(\downarrow\uparrow)}&=&\sqrt{2\mu\left(n_x\hbar\omega_{x}+n_y\hbar\omega_{y}-\epsilon\right)/\hbar^{2}}.
\end{eqnarray}

The scattering amplitudes $f^{(\uparrow\downarrow)}$ and $f^{(\downarrow\uparrow)}$, as well as the coefficients $B_{n_x,n_y}^{(\uparrow\downarrow)}$ and $B_{n_x,n_y}^{(\downarrow\uparrow)}$ are to be determined by solving the Shr\"odinger equation $H|\Psi({\bf r})\rangle=E|\Psi({\bf r})\rangle$. To this end, one can perform the operation $\lim_{\varepsilon\to0}\int_{-\varepsilon}^{+\varepsilon}dz\iint_{-\infty}^{\infty} dx dy \zeta^{1/4}\phi_{n_x}^{*}(\sqrt{\zeta}x)\phi_{n_y}^{*}(y)$ on the both hand sides of the Shr\"odinger equation, then obtains the relations,
\begin{align}
f^{(\uparrow\downarrow)} & =\frac{2\pi|\phi_0(0)|^2\zeta^{1/4}}{ik^{(\uparrow\downarrow)}}\left(a_{s0}\eta^{(\uparrow\downarrow)}+a_{s1}\eta^{(\downarrow\uparrow)}\right),\label{f10}\\
B_{n_x,n_y}^{(\uparrow\downarrow)}&=-\frac{2\pi\zeta^{1/4}}{\kappa_{n_x n_y}^{(\uparrow\downarrow)}}\phi_{n_x}^*(0)\phi_{n_y}^*(0)\left(a_{s0}\eta^{(\uparrow\downarrow)}+a_{s1}\eta^{(\downarrow\uparrow)}\right),\label{f1}\\
f^{(\downarrow\uparrow)} & =\frac{2\pi|\phi_0(0)|^2\zeta^{1/4}}{ik^{(\downarrow\uparrow)}}\left(a_{s0}\eta^{(\downarrow\uparrow)}+a_{s1}\eta^{(\uparrow\downarrow)}\right),\label{f100}\\
B_{n_x,n_y}^{(\downarrow\uparrow)}&=-\frac{2\pi\zeta^{1/4}}{\kappa_{n_x n_y}^{(\downarrow\uparrow)}}\phi_{n_x}^*(0)\phi_{n_y}^*(0)\left(a_{s0}\eta^{(\downarrow\uparrow)}+a_{s1}\eta^{(\uparrow\downarrow)}\right),\label{f2}
\end{align}
where $\eta^{(ss')}= \frac{\partial}{\partial z} \left. \left[z\Psi^{(ss')}(x=0,y=0,z)\right]\right|_{z\to0^{+}}$. Substituting Eqs.~(\ref{f10}-\ref{f2}) into Eqs.~(\ref{psiud}) and (\ref{psidu}) and further into the expression of $\eta^{(ss')}$, one finds the equations for $\eta^{(ss')}$,
\begin{eqnarray}
&&\sqrt{\zeta}\left[\frac{2i}{a_yk^{(\uparrow\downarrow)}}+\Lambda^{(\uparrow\downarrow)}\right]\left(a_{s0}\eta^{(\uparrow\downarrow)}+a_{s1}\eta^{(\downarrow\uparrow)}\right)
\nonumber \\
&&\hspace{3cm}= \frac{\alpha\zeta^{1/4}}{\sqrt{\pi}}-a_y\eta^{(\uparrow\downarrow)},
\label{aerf1}\\
&&\sqrt{\zeta}\left[\frac{2i}{a_yk^{(\downarrow\uparrow)}}+\Lambda^{(\downarrow\uparrow)}\right]\left(a_{s0}\eta^{(\downarrow\uparrow)}+a_{s1}\eta^{(\uparrow\downarrow)}\right)
\nonumber \\
&&\hspace{3cm}=\frac{\beta\zeta^{1/4}}{\sqrt{\pi}}-a_y\eta^{(\downarrow\uparrow)},\label{aerf2}
\end{eqnarray}
where $\Lambda^{(ss')}=\left.\frac{\partial}{\partial z}\left(z\lambda^{(ss')}(z)\right)\right|_{z\to0^{+}}$ with
\begin{align}
\lambda^{(ss')}(z)=\sum_{(n_x,n_y)}'\frac{2a_y \pi}{\kappa_{n_x,n_y}^{(ss')}}|\phi_{n_x}(0)|^2 |\phi_{n_y}(0)|^2 {\rm e}^{-\kappa^{(ss')}_{n_x,n_y}|z|}.
\end{align}

We would like to stress that the derivative $\partial/\partial z$ and the summation over the excited states $\sum^\prime$ can not be interchanged as the summation is not uniformly convergent. Furthermore, it can be proved that in the limit of $z\to0^{+}$, $\lambda^{(ss')}(z)$ can be expanded as
\begin{align}
\label{expansion}
\lambda^{(ss')}(z)=\frac{{\cal C}_{-1}}{z}+{\cal C}+{\cal C}_{1}z+\cdots,
\end{align}
which implies  the existence of $\Lambda^{(ss')}$~\cite{idzia}. The singularity part of $\lambda^{(ss')}(z)$ can be understood in the following way: when the distance of the two colliding atoms approaches zero, the interaction energy will dominate the behavior of the atoms. In this regime, the wave function of the two atoms is not sensitive to the trapping potential, and hence should acquire the same singular behavior as in the three-dimensional geometry. In our calculation, we perform the summation over excited states numerically and fit $\lambda^{(ss')}(z)$ according to the series expansion in Eq.~(\ref{expansion}) to extract the zeroth-order term ${\cal C}$.

\begin{figure}
\centering
\includegraphics[width=0.45\textwidth]{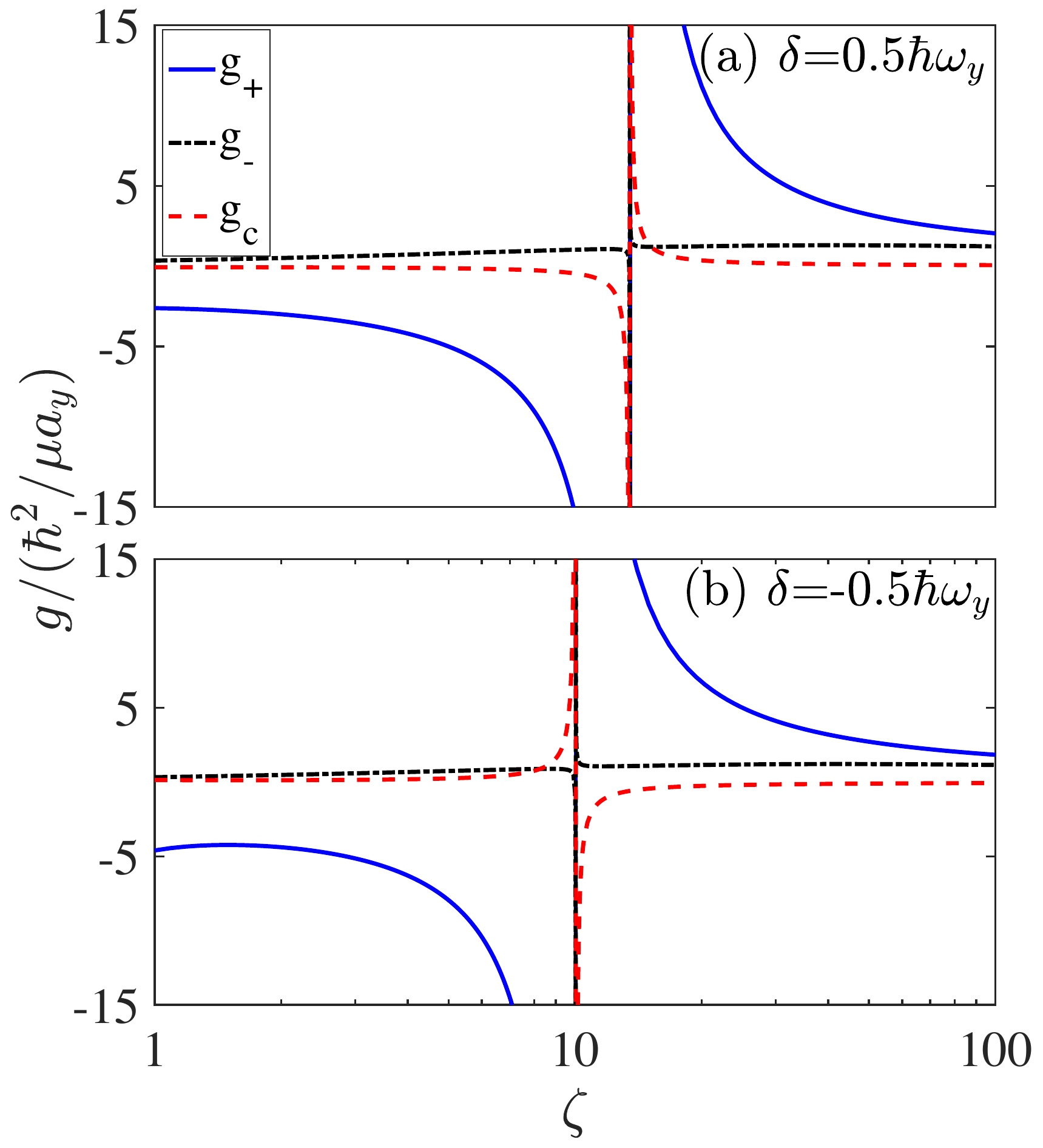}
\caption{Effective one-dimensional interaction $g$'s as functions of anisotropy ratio $\zeta$. Here, we take $\omega_{y}=100$ kHz, and the Zeeman shift as (a) $\delta=0.5\hbar\omega_{y}$ and (b) $\delta=-0.5\hbar\omega_{y}$. Other parameters are same as in Fig.~\ref{g-ay}. In this case, $g_{c}$ is driven to be divergent near the confinement-induced resonance.}\label{g-zeta}
\end{figure}

By solving Eqs.~(\ref{aerf1}) and (\ref{aerf2}) and substituting the expressions for $\eta^{(\uparrow\downarrow)}$ and $\eta^{(\downarrow\uparrow)}$ into Eqs.~(\ref{f10}) and (\ref{f100}), one finally obtains the explicit expression of the scattering amplitudes
\begin{align}
\left(
  \begin{array}{cc}
    f^{(\uparrow\downarrow)}\\
   f^{(\downarrow\uparrow)}
  \end{array}\right)=-(I+iAP)^{-1}
\left(
\begin{array}{cc}
\alpha\\ \beta
\end{array}
\right),\label{quasi1d}
\end{align}
where the parameters $I$, $A$ and $P$ take the form
\begin{align}
I&=\left(\begin{array}{cc}
1 & 0\\
0 & 1
\end{array}\right),\ P=\left(\begin{array}{cc}
k^{(\uparrow\downarrow)} & 0\\
0 & k^{(\downarrow\uparrow)}
\end{array}\right),\label{P}\\
A&=\frac{-a_y^2}{2\sqrt{\zeta}}\left[\left(\begin{array}{cc}
a_{s0} & a_{s1}\\
a_{s1} & a_{s0}
\end{array}\right)^{-1}+\frac{\sqrt{\zeta}}{a_y}\left(\begin{array}{cc}
\Lambda^{(\uparrow\downarrow)} & 0\\
0 & \Lambda^{(\downarrow\uparrow)}
\end{array}\right)\right].\label{am}
\end{align}
It can be seen that the interaction effect is manifested by the presence of scattering lengths $a_{s0}$ and $a_{s1}$ in $A$. We emphasize that Eq.~(\ref{quasi1d}) is valid only in the low-energy limit of $\{\epsilon,\delta\} \ll \{ \hbar \omega_{x}, \hbar \omega_{y} \}$. What one needs to do following is to construct an effective 1D Hamiltonian which can reproduce the scattering amplitude Eq.~(\ref{quasi1d}) in the same energy regime.


\subsection{One-dimension}

\begin{figure*}
\centering
\includegraphics[width=0.85\textwidth]{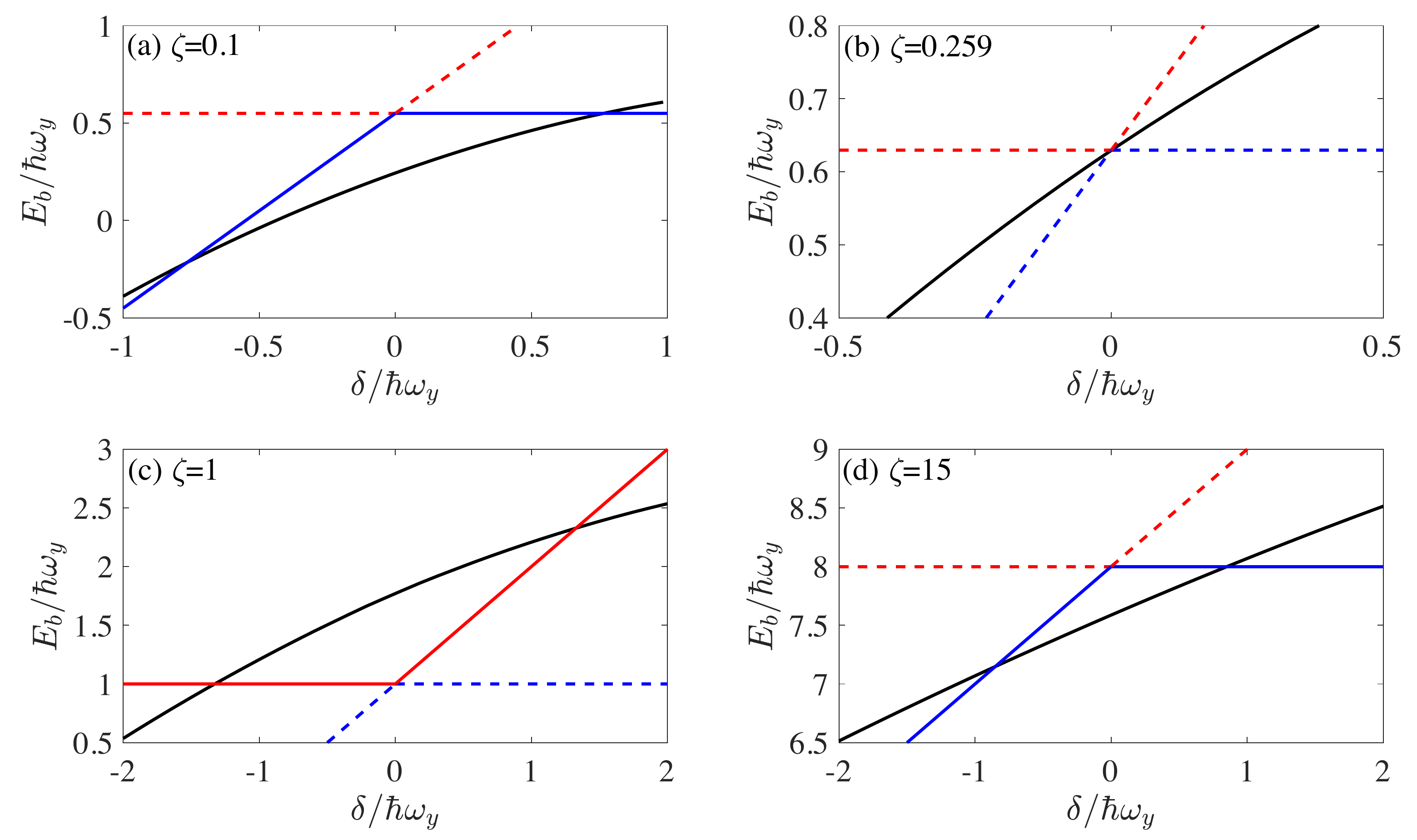}
\caption{Closed channel bound state energy $E_{b}$ (black solid) as a function of $\delta$ for various anisotropy ratios. CIR takes place when the closed channel bound state energy $E_{b}$ is equal to either $E_{\rm th}^{L}$ (blue) or $E_{\rm th}^{H}$ (red). Other parameters are same as in Fig.~\ref{g-zeta}.
\label{Ebfig}}
\end{figure*}
Now we construct an effective 1D model for the quasi-1D Hamiltonian Eq.~(\ref{hamilton}) and analyze the corresponding scattering problem. The interaction part of this effective model should be written as
\begin{align}
\hat{V}_{\rm 1D} & = \left[g_{+}|+\rangle\langle+|+g_{-}|-\rangle\langle-|+g_{c}|+\rangle\langle-|+{\rm h.c.}\right]\delta(z),
\label{v1d}
\end{align}
where $g_{+}$ and $g_{-}$ are the effective 1D interaction strengths corresponding to $V_{+}({\bf r})$ and $V_{-}({\bf r})$, respectively, and $g_{c}$ denotes the coupling between the $|+\rangle$ and $|-\rangle$ channels. We would like to stress that  unlike the quasi-1D model, the effective interaction is in general not diagonal in the 1D case even in the basis of $\{|+\rangle,|-\rangle\}$. The underlying reason is that in the presence of magnetic field, the free Hamiltonian is non-diagonal in the basis of $\{|+\rangle,|-\rangle\}$. When the excited states are integrated out, the coupling effect between $|+\rangle$ and $|-\rangle$ induced by the free Hamiltonian can also be manifested in the effective interaction, leading to a nonzero $g_{c}$. However, in the absence of magnetic field, the $|+\rangle$ and $|-\rangle$ channels decouple, and the quasi-1D Hamiltonian Eq.~(\ref{hamilton}) can be reduced to two separate single-channel models with their corresponding interactions $g_\pm$. CIR within each channel then takes place at the position given by the single-channel calculation~\cite{Shiguo-Hui,Wei-Peng,schmelcher,saenz}.  A similar effect has been discussed in the exploration of CIR in spinor atoms.~\cite{XL-spinor}

In the basis of $\{|g\uparrow;e\downarrow\rangle,|g\downarrow;e\uparrow\rangle\}$, the effective 1D Hamiltonian can be written as
\begin{align}
\hat{H}_{\rm 1D} & = \left(-\frac{\hbar^{2}}{2\mu}\frac{d^{2}}{dz^{2}}\right)\left({\cal P}_{\uparrow\downarrow}+{\cal P}_{\downarrow\uparrow}\right)+\delta{\cal P}_{\uparrow\downarrow}\nonumber\\&+\Bigg[\frac{g_{+}+g_{-}+2g_{c}}{2}{\cal P}_{\uparrow\downarrow}+\frac{g_{+}+g_{-}-2g_{c}}{2}{\cal P}_{\downarrow\uparrow}\nonumber\\&+\frac{g_{-}-g_{+}}{2}({\cal S}_{{\rm ex}}+{\cal S}_{{\rm ex}}^{\dagger})\Bigg]\delta(z),
\label{h1d}
\end{align}
where ${\cal P}_{ss'}$ and ${\cal S}_{\rm ex}$ have the same definition as that in Eq.(\ref{hamilton}). Similar to the case of quasi-1D system, the general scattering wave function of Eq.~(\ref{h1d}) can be written as
\begin{align}
|\Psi_{{\rm 1D}}(z)\rangle&=\left(\alpha e^{ik^{(\uparrow\downarrow)}z}+f_{{\rm 1D}}^{(\uparrow\downarrow)}e^{ik^{(\uparrow\downarrow)}|z|}\right)|g\uparrow;e\downarrow\rangle\nonumber\\&+\left(\beta e^{ik^{(\downarrow\uparrow)}z}+f_{{\rm 1{\rm D}}}^{(\downarrow\uparrow)}e^{ik^{(\downarrow\uparrow)}|z|}\right)|g\downarrow;e\uparrow\rangle.\label{psi1d}
\end{align}
After some straightforward calculation, one can  obtain the scattering amplitudes $f_{{\rm 1D}}^{(\uparrow\downarrow)}$
and $f_{{\rm 1{\rm D}}}^{(\downarrow\uparrow)}$ of the 1D model,
\begin{align}
\left(\begin{array}{c}
f_{{\rm 1D}}^{(\uparrow\downarrow)}\\
f_{{\rm 1D}}^{(\downarrow\uparrow)}
\end{array}\right)=-(I+iA_{{\rm 1D}}P)^{-1}\left(\begin{array}{c}
\alpha\\
\beta
\end{array}\right),\label{fr-1-1}
\end{align}
where $I$ and $P$ are given in Eq.~(\ref{P}), and $A_{\rm 1D}$ is defined as
\begin{equation}
A_{{\rm 1D}}=-\frac{2\hbar^{2}}{\mu}\left(\begin{array}{cc}
g_{+}+g_{-}+2g_{c} & g_{-}-g_{+}\\
g_{-}-g_{+} & g_{+}+g_{-}-2g_{c}
\end{array}\right)^{-1}.\label{a1d}
\end{equation}
By comparing  Eq.~(\ref{a1d}) with Eq.~(\ref{quasi1d}), and requiring $f_{{\rm 1D}}^{(\uparrow\downarrow)}=f^{(\uparrow\downarrow)}$ and $f_{{\rm 1D}}^{(\downarrow\uparrow)}=f^{(\downarrow\uparrow)}$, one finally obtains the equation of the effective 1D interaction strengths
\begin{widetext}
\begin{align}
\left(\begin{array}{cc}
a_{s0} & a_{s1}\\
a_{s1} & a_{s0}
\end{array}\right)^{-1}+\frac{\sqrt{\zeta}}{a_{y}}\left(\begin{array}{cc}
\Lambda^{(\uparrow\downarrow)} & 0\\
0 & \Lambda^{(\downarrow\uparrow)}
\end{array}\right)=\frac{4\hbar^{2}\sqrt{\zeta}}{\mu a_{y}^{2}}\left(\begin{array}{cc}
g_{+}+g_{-}+2g_{c} & g_{-}-g_{+}\\
g_{-}-g_{+} & g_{+}+g_{-}-2g_{c}
\end{array}\right)^{-1}.\label{g1d}
\end{align}
\end{widetext}
For the particular case of $\zeta=1$, $\Lambda^{(\uparrow\downarrow)}$ and $\Lambda^{(\downarrow\uparrow)}$ can be simplified as $\zeta_{H}(1/2,1+\delta/\hbar\omega_{y})$ and $\zeta_{H}(1/2,1)$, respectively, where $\zeta_{H}(a,d)$ denotes the Hurwitz-Zeta function. Thus, Eq.~(\ref{g1d}) reproduces the result in Ref.~\cite{ren}.
For another particular case that the magnetic field is zero, the $|+\rangle$ and $|-\rangle$ channels decouple and the solution of Eq.~(\ref{g1d}) can be significantly simplified as
\begin{eqnarray}
g_{\pm}=\frac{2\hbar^{2}\sqrt{\zeta}}{\mu a_{y}\left(a_{y}/a_{s\pm}+\sqrt{\zeta}\Lambda\right)},\quad g_c=0, \label{deltaeq0}
\end{eqnarray}
where $\Lambda\equiv\Lambda^{(\uparrow\downarrow)}=\Lambda^{(\downarrow\uparrow)}$. Furthermore, it can be proved that in the isotropic case of $\zeta=1$, $\Lambda={\cal C}=-\zeta_{H}(1/2,1)$, which recovers the result of a single-channel model~\cite{oshanii98}.

\section{Effective interaction}
\label{sec:interaction}

For the general case with transversal anisotropy and a finite magnetic field, one can obtain the effective 1D interaction strengths by solving Eq.~(\ref{g1d}) numerically. In Fig.~\ref{g-ay}, we present the variation of the effective interaction strengths $g_\pm$ and $g_c$ by varying the transverse trapping potential for a set of different anisotropy ratios with Zeeman shift $\delta=0$. In this figure, the transverse trap is characterized by the length scale $a_y \equiv \sqrt{\hbar /(\mu \omega_y)}$ associated with the $y$-axis harmonic potential. In each panel, the resonance at the lower end of $a_{y}/a_{s-}$ is associated with the potential $V_{-}({\bf r})$, while the one at the higher end of $a_{y}/a_{s-}$ is associated with $V_{+}({\bf r})$. Specifically, in Fig.~\ref{g-ay}(a), the system is isotropic in the transverse plane with $\zeta=1$ and the locations of the two resonances recover the result of Olshanii with $a_{y}/a_{s-}=-\zeta_{H}(1/2,1)$ and $a_{y}/a_{s+}=-\zeta_{H}(1/2,1)$, respectively. In Fig.~\ref{g-ay}(b), we plot the case of $\zeta=5$, where the locations of the two resonances are shifted to the right and reach $a_{y}/a_{s-}=1.6$ and $a_{y}/a_{s+}=1.6$, respectively. If the anisotropy is further enhanced, the locations of CIRs move back to those of the isotropic case at about $\zeta=6.5$ as shown in Fig.~\ref{g-ay}(c), and then shift to the further left and reach $a_{y}/a_{s-}=1.0$ and $a_{y}/a_{s+}=1.0$ at $\zeta=10$. This observation shows that in the aid of transversal anisotropy, the location of CIR can be tuned to a large extent, which is facilitative for the control of CIR in AE atoms in experiments.

In Fig.~\ref{g-zeta}, we show the effective 1D interactions as functions of anisotropy ratio $\zeta$ for various magnetic fields. In this calculation, we take $\omega_{y}=100$ kHz and increase $\omega_{x}$ to vary the anisotropy ratio $\zeta$. As the Zeeman field couples both the $|+\rangle$ and $|-\rangle$ channels, the interactions $g_\pm$ and $g_c$ all become divergent at CIR. In Fig.~\ref{g-zeta}(a) and \ref{g-zeta}(b), we show the CIR associated with the $|-\rangle$ channel, while the Zeeman shifts are taken as $\delta=0.5\hbar\omega_y$ and $\delta=-0.5\hbar\omega_y$, respectively. First of all, it is clear that the location of CIR can be changed by tuning the magnetic field, which works together with the anisotropy ratio to provide a versatile toolbox to tune CIR to experimentally favorable regimes. Secondly, $g_{c}$ is driven to be divergent near CIR. This phenomenon implies that the effective 1D interaction no longer respects to the nuclear spin rotational symmetry and couples the $|+\rangle$ and $|-\rangle$ channels by containing non-zero $g_{c}$.

\section{Bound state analysis}
\label{sec:boundstate}

CIR can also be understood as a Feshbach resonance. The ground state of the transversal trapping potential with $n_x=n_y=0$ is considered as the open channel, while all the excited states as a whole compose the closed channel. The resonance takes place when the bound state energy of the closed channel becomes degenerate with the open channel threshold.

The bound state energy $E_b$ of the closed channel can be obtained by solving the Shr\"odinger equation ${\cal P}_c^\dagger \hat{H} {\cal P}_c |\Psi_{b}\rangle = E_{b}|\Psi_{b}\rangle$ with $\hat{H}=\hat{H}_{\rm Q1D}+\mu \omega_{z}^2z^2/2$ and ${\cal P}_c$ the projection operator associated with the closed channel~\cite{Wei-Peng,idzia}, and then take the limit $\omega_{z}\rightarrow 0$. The resulting equation reads
\begin{align}
\left(\begin{array}{cc}
a_{s0} & a_{s1}\\
a_{s1} & a_{s0}
\end{array}\right)^{-1}+\frac{\sqrt{\zeta}}{a_{y}}\left(\begin{array}{cc}
\Lambda^{(\uparrow\downarrow)} & 0\\
0 & \Lambda^{(\downarrow\uparrow)}
\end{array}\right)=0,\label{boundstateE}
\end{align}
where $\Lambda^{({\uparrow\downarrow})}={\cal F}(E_{b}+\delta)/\sqrt{\pi\zeta}$ and $\Lambda^{({\downarrow\uparrow})}={\cal F}(E_{b})/\sqrt{\pi\zeta}$ with
\begin{align}
F({\cal E})=\int_{0}^\infty dt\left(\frac{\sqrt{\zeta}e^{-{\cal E}t}}{\sqrt{(1-e^{-\zeta t})(1-e^{-t})t}}-\frac{\sqrt{\zeta}te^{-{\cal E}t}+1}{t^{3/2}}\right).
\end{align}
Notice that this expression is consistent with Eq.~(\ref{g1d}) by setting the right-hand-side to zero. The reason for this consistency is that when CIR takes place, i.e., $g's\to \infty$, there always exit a bound state near the threshold energy.

\begin{figure}
\centering
\includegraphics[width=0.45\textwidth]{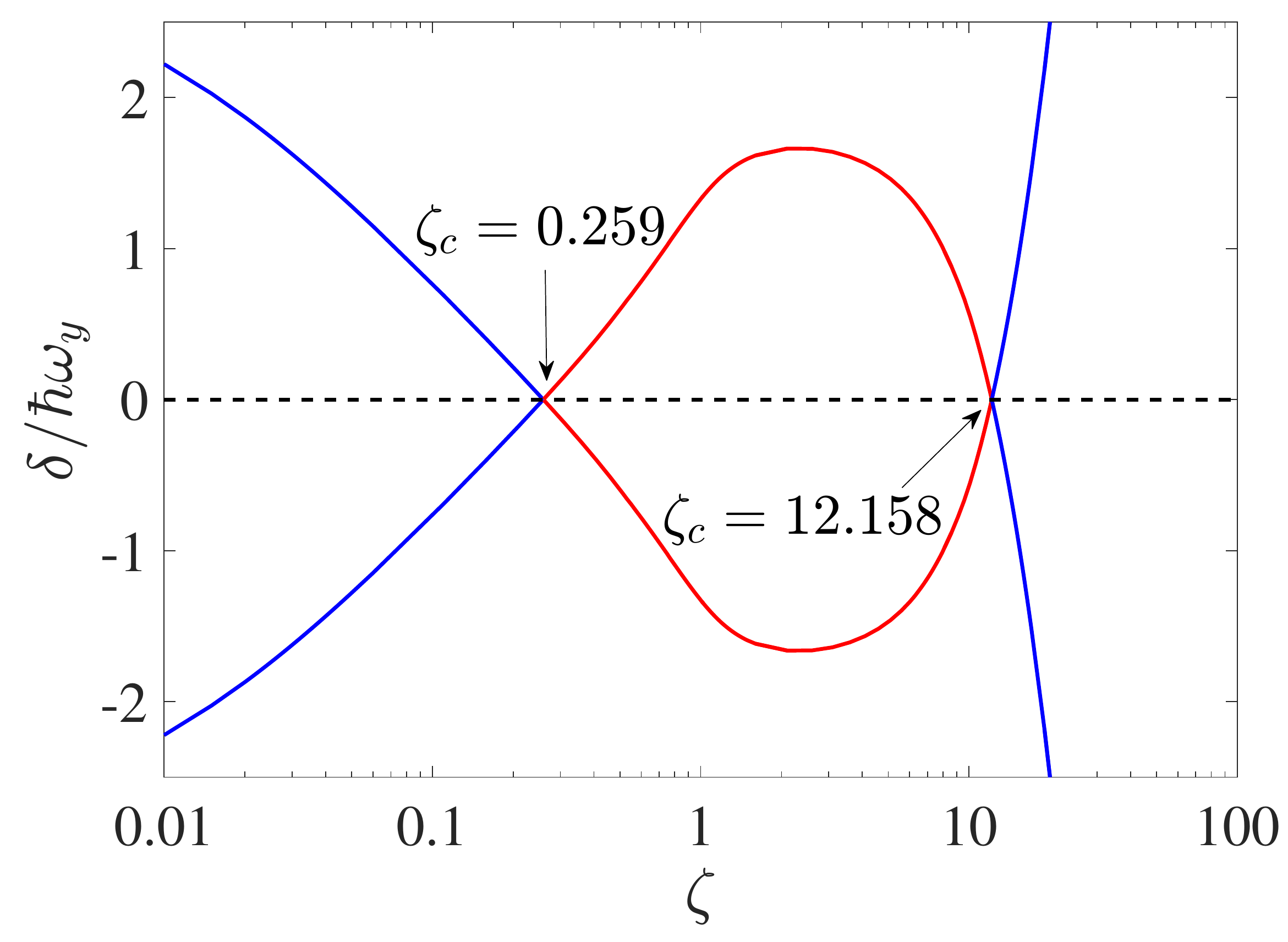}
\caption{Locations of CIR in the parameter plane spanned by $\zeta$ and $\delta$. In the regions of $\zeta \lesssim 0.259$ and $\zeta \gtrsim 12.158$ (blue), CIR takes place when the closed channel bound state energy intersects with the lower threshold $E_{\rm th}^{L}$, while in the region of $0.259 \lesssim \zeta \lesssim 12.158$ (red), CIR happens when the closed channel bound state intersects with the higher threshold $E_{\rm th}^{H}$. Other parameters are same as in Fig.~\ref{g-zeta}.
\label{delta-zeta}}
\end{figure}

By solving Eq.~(\ref{boundstateE}), one obtains the closed channel bound state energy as shown in the insets of Fig.~
\ref{g-ay} for $\delta = 0$ and in Fig.~\ref{Ebfig} for the case of finite $\delta$. Notice that in all cases the intersection between the closed channel bound state energy and the open channel threshold takes place at the same position of diverging effective 1D interaction, i.e., the location of CIR. This concludes that the results obtained by solving a scattering problem are consist with those of the bound state analysis.

As in the vicinity of an OFR, the Zeeman energies of the $|g\uparrow;e\downarrow\rangle$ and $|g\downarrow;e\uparrow\rangle$ channels are detuned by $\delta$, one can respectively define a threshold energy for each of them, i.e., $E_{\rm th}^{(\uparrow\downarrow)}=\hbar(\omega_{x}+\omega_{y})/2+\delta$ and $E_{\rm th}^{(\downarrow\uparrow)}=\hbar(\omega_{x}+\omega_{y})/2$. For convenience, we can further define the lower threshold and the higher threshold energy as $E_{\rm th}^{L}=\min(E_{\rm th}^{(\uparrow\downarrow)},E_{\rm th}^{(\downarrow\uparrow)})$ and $E_{\rm th}^{H}=\max(E_{\rm th}^{(\uparrow\downarrow)},E_{\rm th}^{(\downarrow\uparrow)})$, denoted by the blue and red lines in Fig.~\ref{Ebfig}, respectively. Notice that the difference between $E_{\rm th}^{L}$ and $E_{\rm th}^{H}$, which is the Zeeman energy shift $\delta$, is in general much smaller than the trapping frequency of the transversal confinement. Thus, both of the two scattering channels are relevant to the quasi-1D low-energy scattering process.

In Fig.~\ref{Ebfig}, we show the closed channel bound state energy as a function of Zeeman energy for various anisotropy ratios. Our results show that in the regions of $\zeta \lesssim 0.259$ and $\zeta \gtrsim 12.158$, CIR takes place when the closed channel bound state degenerates with the lower threshold energy with $E_{b}=E_{\rm th}^{L}$, while in the region of $0.259 \lesssim \zeta \lesssim 12.158$, CIR happens when the closed channel bound state intersects with the higher threshold energy with $E_{b}=E_{\rm th}^{H}$. At the critical points $\zeta_{c}=0.259$ and $12.158$, CIR is located at the position of zero Zeeman energy difference $\delta = 0$. We also notice that the positions of CIR are symmetric with respect to $\delta=0$, which reflects the symmetry between the $|g\uparrow;e\downarrow\rangle$ and $|g\downarrow;e\uparrow\rangle$ channels with opposite Zeeman energy shift. The two branches of CIR with varying anisotropy ratio and the Zeeman energy detuning are summarized in Fig.~\ref{delta-zeta}.

\section{conclusion}
\label{sec:conclusion}

In summary, we study the two-body problem of alkaline-earth-metal-like atoms confined in quasi-one-dimensional tube with transversal anisotropy. By solving the scattering problem, we find that the anisotropy ratio of the trapping potential can serve as a knob to tune the location of confinement-induced resonance. This extra controllability can work together with the Zeeman energy to facilitate the tuning of CIR in experiments. Then we explore the CIR from the viewpoint of bound state. Since in the vicinity of an orbital Feshbach resonance, the detuning of the two channels are in general much smaller than the transversal trapping frequency, both scattering channels are of relevance in low-energy physics, such that two open channel threshold energies $E_{\rm th}^{L}$ and $E_{\rm th}^{H}$ can be defined. We find that CIR takes place when the closed channel bound state degenerates to either $E_{\rm th}^{L}$ or $E_{\rm th}^{H}$. Our results can be readily checked in current experiments, where the position of CIR can be determined by measuring the atom-loss rate and the bound state binding energy can be extracted from the radio frequency spectrum.

{\it Acknowledgement. }
This work is supported by the National Key R\&D Program of China (Grant No. 2018YFA0306501(W.Z.), 2018YFA0307601 (R.Z.)), the National Natural Science Foundation of China (Grant Nos. 11434011, 11522436, 11704408, 11774425, 11804268 (RZ)), and the Research Funds of Renmin University of China (Grant Nos. 16XNLQ03, 18XNLQ15). X.Z. acknowledges support from the National Postdoctoral Program for Innovative Talents (Grant No. BX201601908) and the China Postdoctoral Science Foundation (Grant No. 2017M620991).

\end{document}